\documentclass[aps,prl,twocolumn,superscriptaddress,showpacs, nofootinbib]{revtex4-2}
\usepackage{flushend}
\usepackage{balance}
\usepackage{amssymb}
\usepackage{suffix}
\usepackage{mathtools}
\usepackage[utf8]{inputenc}
\usepackage{booktabs}
\usepackage{cases}
\usepackage[multiple]{footmisc}
\usepackage{dcolumn}
\usepackage{color,soul}
\usepackage{rotating}
\usepackage{perpage}
\usepackage{siunitx}
\usepackage{xcolor}
\usepackage{soul}
\usepackage{amsmath}
\usepackage{dsfont}
\usepackage{tikz}
\usepackage[T1]{fontenc}
\usepackage{etoolbox}
\usepackage{graphics}
\usepackage{siunitx}
\usepackage[hidelinks]{hyperref}
\hypersetup{
    colorlinks,
    linkcolor={red},
    citecolor={blue},
    urlcolor={blue}
}
\usepackage{float}	
\usepackage{collref}
\usepackage{multirow}
\usepackage{mathtools}
\usepackage{bm}
\usepackage{url}

\usepackage{tikz}
\usepackage{tikz-3dplot}

\usepackage{xcolor}
\newcommand{\blue}[1]{\textcolor{black}{#1}}
\usepackage{comment}
\usepackage{braket}
\definecolor{cobalt}{rgb}{0.0, 0.28, 0.67}
\newcommand{\re}{\textcolor{purple}}

\begin{document}

\title{Topological anomalous Floquet photon pump}

\author{Manshuo Lin}
\email{manshuo.lin@utdallas.edu}
\address{Department of Physics, The University of Texas at Dallas, Richardson, Texas 75080, USA}

\author{Saeed Rahmanian Koshkaki}
\email{rahmanian@tamu.edu}
\address{Department of Physics, The University of Texas at Dallas, Richardson, Texas 75080, USA}
\address{Department of Chemistry, Texas A \& M University, College Station, Texas 77842, USA}

\author{Mohsen Yarmohammadi}
\email{mohsen.yarmohammadi@georgetown.edu}
\address{Department of Physics, The University of Texas at Dallas, Richardson, Texas 75080, USA}
\address{Department of Physics, Georgetown University, Washington DC 20057, USA}

\author{Michael H. Kolodrubetz}
\email{mkolodru@utdallas.edu}
\affiliation{Department of Physics, The University of Texas at Dallas, Richardson, Texas 75080, USA}

\date{\today}

	\begin{abstract}
There is a close theoretical connection between topological Floquet physics and cavity QED, yet this connection has not been realized experimentally due to complicated cavity QED models that often arise. We propose a simple, experimentally viable protocol to realize non-adiabatic topological photon pumping mediated by a single qubit, which we dub the anomalous Floquet photon pump. For both quantized photons and external drive, the system exhibits a non-trivial topological phase across a broad range of parameter space. Transitions out of the topological phase result from frequency-space delocalization. Finally, we argue that the protocol can be implemented in existing experiments via driven qubit non-linearities, with topological pumping witnessed in measurements of the cavity Wigner distribution functions.
 \end{abstract}
	\maketitle

	{\allowdisplaybreaks	

\begin{figure}[t]
		\centering
		\includegraphics[width=1\linewidth]{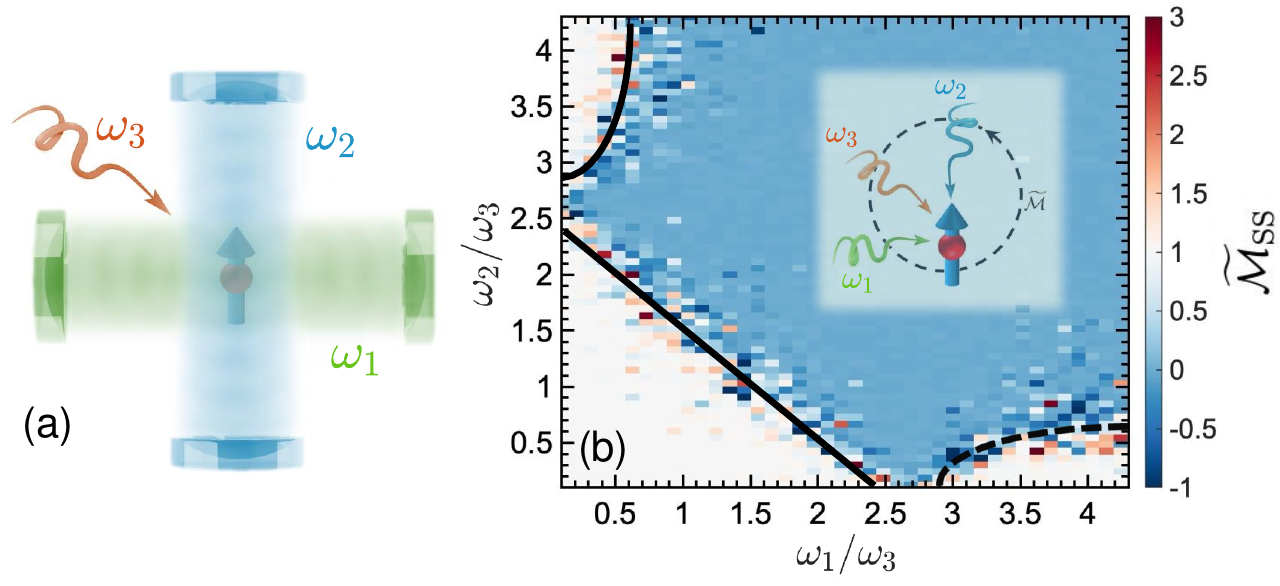}
		\caption{
\textbf{Illustration of qubit-mediated topological photon pump.} Schematics of (a) the anomalous Floquet photon pump (AFPP) and (b, inset) the anomalous Floquet qubit pump (AFQP) show the qubit interacting with three incommensurate drives. (b) Phase diagram for late-time average magnetization $\widetilde{\mathcal M}_{\rm SS}$ in the AFQP. The black dashed lines represent approximate transition lines between trivial and non-trivial phases.}
		\label{phase_diagram_AFQP}
\end{figure}


 \textbf{\blue{\textit{Introduction.}}}--- A canonical example of topological physics is the Thouless pump, which uses adiabatic tuning of parameters and a filled topological band to enable quantized charged pumping protected by a Chern number \cite{Thouless_quantized_particle_transport, Thouless_quantized_hall_2d_periodic, Thouless_pumping}. The nature of topological protection suggests that quantized pumping will hold with incommensurate potentials~\cite{Thouless_quantized_particle_transport}, substrate effects, or many-body interactions~\cite{Thouless_qunatized_adiabatic_charge_transport}. However, if the system operates at finite frequencies (non-adiabatic regime), the quantized transport does not generally survive~\cite{Shih_nonadiabatic_partical_transport, Niu_Qpump_charges, Privitera_nonadiabatic_breaking_TP, Anushya_absence_of_disorderedTP}. 
While quantized pumping can be restored by considering the Floquet Hamiltonian in certain limits~\cite{Niu_Qpump_charges, Savvas_rapid_cycle_TP},
it is much less robust than the adiabatic pump.

By contrast, a distinct topological phase of matter can exist in a driven disordered 2D system, known as the anomalous Floquet-Anderson insulator (AFAI), in which topologically protected quantized charge pumping occurs due to chiral edge states ~\cite{Rafael_AFAI_nonadiabatic_qcp, Rudner_Anomalous_BE_correspondence}. 
By dimensional reduction, similar topological phases can be obtained in 1D with two incommensurate drives or in 0D (a few-level system) coupled to three incommensurate drives \cite{Long_Nonadiabatic, NathanPRLqFTEP}. 
This family of topological pumps is known as the quasiperiodic Floquet-Thouless energy pump or anomalous localized topological phase. Crucially, it is stabilized by localization -- both spatial (Anderson) localization and frequency-space (Wannier-Stark) localization -- as well as non-adiabatic drives~\cite{Shirley_Floquet, NathanPRLqFTEP}. Therefore, these phases give rise to topological phenomena that are fundamentally more stable than their more well-known cousins, such as the Chern insulator and Thouless pump~\cite{Timms_QFT_Noise}. 

The AFAI has been realized in engineered quantum systems such as cold atoms in optical lattices~\cite{ultracold_Flo_topo, Maczewsky_photonic_anomalous_FI, Li_Fractal_PAFI}
, while its 1D or 0D variants have not been experimentally realized.  
Suggested routes to realizing the 0+3D system involve using a single qubit driven with three incommensurate frequencies~\cite{Long_Nonadiabatic, NathanPRLqFTEP}, but realizations proposed to date involved high-harmonic and multi-photon driving that is not viable in current experiments. 

In this Letter, we introduce a much simpler model of topological
photon pump which should be accessible in current experiments. We show that our model exhibits a topologically non-trivial phase in which topological pumping occurs non-adiabatically between the photonic degrees of freedom. The phase diagram depends on whether the photon is treated quantum mechanically or as an external drive; in the former case, finite photon frequency is shown to be crucial for the stability of the topological phase. Finally, we propose experimentally relevant observables to detect the topological pumping, such as the Wigner distribution function of the cavity modes \cite{Wigner_on_QC_for_TE, Kurtsiefer_Wigner_helium, Lutterbach_Wigner_CavityQED_IonTrap, Bertet_Wigner_Fock_Cavity, Banaszek_Wigner_counting}.

\textbf{\blue{\textit{AFQP Model.}}}--- The model we consider throughout this Letter consists of a single qubit driven by three drives at incommensurate frequencies, as represented in Fig.~\ref{phase_diagram_AFQP}(a). We start by introducing the simplest version of this problem, in which all three drives are treated as external (classical) modulations. 
The system evolves in Floquet-like cycles with frequency $\omega_3=2\pi/T_3$ and time-dependent Hamiltonian ($m \in \mathds Z$):
\begin{subequations}\label{H_t_afqp}
\begin{align}
    H(t) & = H_j(t) \quad \text{for} \quad m+\tfrac{j-1}{3} < t/T_3 < m+\tfrac{j}{3}\, \nonumber,\\
    H_1(t) & = -J_0(e^{-i\omega_1 t}\sigma^+ + e^{i\omega_1 t}\sigma^-)\,, \\
    H_2(t) & = -J_0(e^{i\omega_2 t}\sigma^+ + e^{-i\omega_2 t}\sigma^-)\,, \\ 
    H_3(t) & = -J_0\sigma^x\,.   
\end{align}
\end{subequations}
We adopt units where $\omega_3=\hbar=1$ and choose the coupling strength $J_0 \equiv 3\,\omega_3/4$ so that the qubit undergoes an exact $\pi$-rotation during each interval $T_3/3$  if  $\omega_{1,2}\to 0$. Beyond this limit, $\omega_{1,2}$ breaks the drive's periodicity. In particular, frequencies are chosen to be incommensurate such that $\omega_a/\omega_b$ is irrational for any $a\neq b$. As this can be viewed as a generalized quasiperiodic Floquet system with three classical Floquet drives, we call it the anomalous Floquet qubit pump (AFQP).


The topological response of the AFQP is a quantized photonic ``magnetization'' 
in which energy is transferred cyclically between the drives $\omega_{1,2,3}$ analogous to a current loop. Explicitly, the topological winding number $W$ of the generalized Floquet micromotion operator is equal to the long-time average of magnetization, $W \approx \widetilde{\mathcal M}_{\rm SS}$ \cite{Long_Nonadiabatic}, where 
\begin{equation}
 \small  \widetilde{\mathcal M}(t)  = \frac{\pi}{2|\omega| T_3} \int_{t}^{t+T_3} dt^\prime \,  {\rm{Tr}_{\rm spin}}\left[ \left(\dot{\vec{n}}(t^\prime) \times \vec{n} (t^\prime)\right) \cdot \hat{\omega} + {\rm h.c.} \right] \, ,
\end{equation}
with $\vec \omega = (\omega_1,\omega_2,\omega_3)$. To probe the topological properties of the AFQP, we numerically compute the late time average of $\widetilde{\mathcal M}$ between $t=8\times 10^3~T_3$ and $10^4~T_3$, which we denote $\widetilde{\mathcal M}_{\rm SS}$. Irrational frequency ratios are maintained by setting $\omega_2/\omega_3= p\sqrt{2}/1000$ and $\omega_1/\omega_3= k\sqrt{2}/1000\sqrt{3}$ for $\{k,p\} \in \mathds{Z}$. 
For the AFQP, in which drives are not quantized, we instead treat the photons semi-classically through the (qubit space) photon current operator $\dot{n}_j(t) \equiv -U^\dag \left(\partial_{\omega_j t} H \right) U $ where $U(t)$ is the time-evolution unitary and  $  j\in \{1,2,3\}$.  

The topological phase diagram of the AFQP is shown in Fig.~\ref{phase_diagram_AFQP}(b). A clear topological phase ($W=1$, white regions)  occurs across a wide range of drive frequencies. The phase transition to a topologically trivial phase occurs due to the breakdown of frequency localization, similar to the phase diagram found in previous models \cite{NathanPRLqFTEP, Long_Nonadiabatic}. However, our model for the AFQP is far simpler and more experimentally feasible; indeed, this AFQP magnetization can readily be measured in most  modern qubit setups using appropriate time-dependent control.
\begin{figure}[t]
		\centering
        \includegraphics[width=1\linewidth]{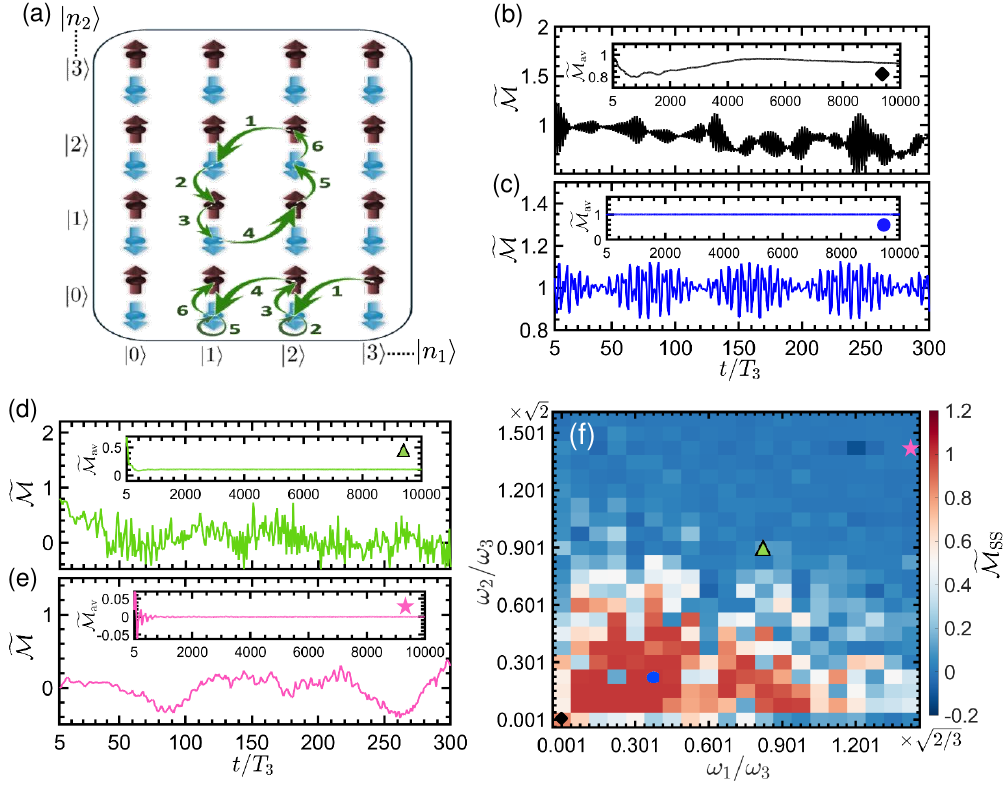}
		\caption{\textbf{Topological phase diagram of the anomalous Floquet photon pump (AFPP).} (a) Frequency lattice illustration of photon pumping in the limit $\omega_{1,2} \ll \omega_3$, showing both bulk magnetization and edge current. The lattice sites represent different spin and photon Fock states $\vert S,n_1,n_2\rangle$. (b-e) Time dependence of $\widetilde{\mathcal M}$ at various frequencies. Insets show running average of $\widetilde{\mathcal M}$ between time $0$ and $t$. (b) For $\omega_{1,2}\ll \omega_3$, $\widetilde{\mathcal M}$ deviates notably from $1$. (c) For larger $\omega_{1,2}\sim 0.3\,\omega_3$, quantization is restored. (d) For $\omega_{1,2}\sim \omega_3$, $\widetilde{\mathcal M}$ converges to a non-quantized value. (e) For $\omega_{1,2}\gg \omega_3$, $\widetilde{\mathcal M}$ converges to $0$. (f) Full phase diagram for the AFPP from late-time magnetization, $\widetilde{\mathcal M}_\mathrm{SS}$, calculated as an average of $\widetilde{\mathcal M}$ from $8\times 10^3~T_3$ to $10^4~T_3$
  . All data shown have photon cutoffs $n_{c1,c2}=19$, initial occupation $\mathcal{N}_{i1,i2}=9$, coupling constants $J_{j}=3/(4\sqrt{\mathcal{N}_{ij}})=0.25$ for $j=\{1,2\}$. }
		\label{phase_diagram_AFPP}
\end{figure}

\textbf{\blue{\textit{AFPP Model.}}}--- While the AFQP is easy to realize, the observable magnetization $\widetilde{\mathcal M}$ is somewhat artificial. A more physical interpretation comes from treating the drives $\omega_{1,2}$ as quantized photons, which is mathematically obtained via Fourier transforming and introducing an extended Hilbert space \cite{NathanPRLqFTEP}, which is physically interpreted as replacing $e^{i\omega_j t} \to a_j$ due to Heisenberg evolution of the cavity mode~\cite{Long_Nonadiabatic}. This results in a model where the qubit couples to two quantized cavity modes, $\omega_1$ and $\omega_2$, and one classical drive, $\omega_3$: 
\begin{subequations}\label{H_t_afpp}
    \begin{align}
        H_1 &= -J_1(\sigma^+  a_1^\dag+ \sigma^-  a_1) + \omega_1  n_1  +\omega_2  n_2 \,,\\
        H_2 &= -J_2(\sigma^+ a_2 + \sigma^- a_2^\dag  ) + \omega_1  n_1  +\omega_2  n_2 \,,\\
        H_3 &= -J_0\sigma^x  + \omega_1  n_1  +\omega_2  n_2 \,.
    \end{align}
\end{subequations}
We call this model the anomalous Floquet photon pump (AFPP). Unlike previous models, the terms in $H_{1,2,3}$ are canonical Hamiltonians from cavity QED with Jaynes-Cummings (JC) ~\cite{Jaynes_cumming}, anti-JC, or driven qubit form. In order to connect its high-photon-number limit to the semi-classical AFQP, coupling constants are scaled as $J_j = J_0 / \sqrt{N_{i,j}}$, where $N_{i,j}$ is the non-zero initial photon number in mode $j \in \{1,2\}$. If the initial photon number of mode $j$ is zero, the corresponding $N_{i,j}$ will be set to $1$. 
Topological pumping is readily seen in the limit $\omega_{1,2} \ll \omega_3$, as the photons undergo a ``magnetization'' loop enclosing 2 unit cells in 2 Floquet cycles -- see Fig.~\ref{phase_diagram_AFPP}(a). Furthermore, the natural photonic ``edge'' at $n=0$ implies the existence of topological edge states, which pump energy to the remaining photon at a quantized rate, as also shown in Fig.~\ref{phase_diagram_AFPP}(a).

Although Eqs.~\eqref{H_t_afpp} define infinite-dimensional Hamiltonians in a frequency lattice, in practice, we use finite photon cutoffs $N_{c1,c2}$ for simulations and initialize the photons in Fock states at approximately $N_c/2$ unless stated otherwise. We determine the topological phase diagram by time-evolving this initial state and calculating the average of $\widetilde{\mathcal{M}}$ at late times. The operators $n_{1,2}$ and $\dot n_{1,2}$ now refer to real photons, while the trace over spins states is replaced by a sum over expectation values starting the spin states up and down. As seen in Fig.~\ref{phase_diagram_AFPP}(b)-(e), both topologically trivial (e) and non-trivial (c) points remain with quantized magnetization $\widetilde{\mathcal{M}}_{\mathrm{SS}}$. However, at low (b) or intermediate (d) frequencies, the signal does not appear quantized, consistent with an irregular topological transition involving photon-space delocalization and a possible intervening delocalized phase, as seen in~\cite{NathanPRLqFTEP, Long_Nonadiabatic}
. This is clearly seen in the phase diagram as well, Fig.~\ref{phase_diagram_AFPP}(f).

\begin{figure}[t]
	\centering
  \includegraphics[width=1\linewidth]{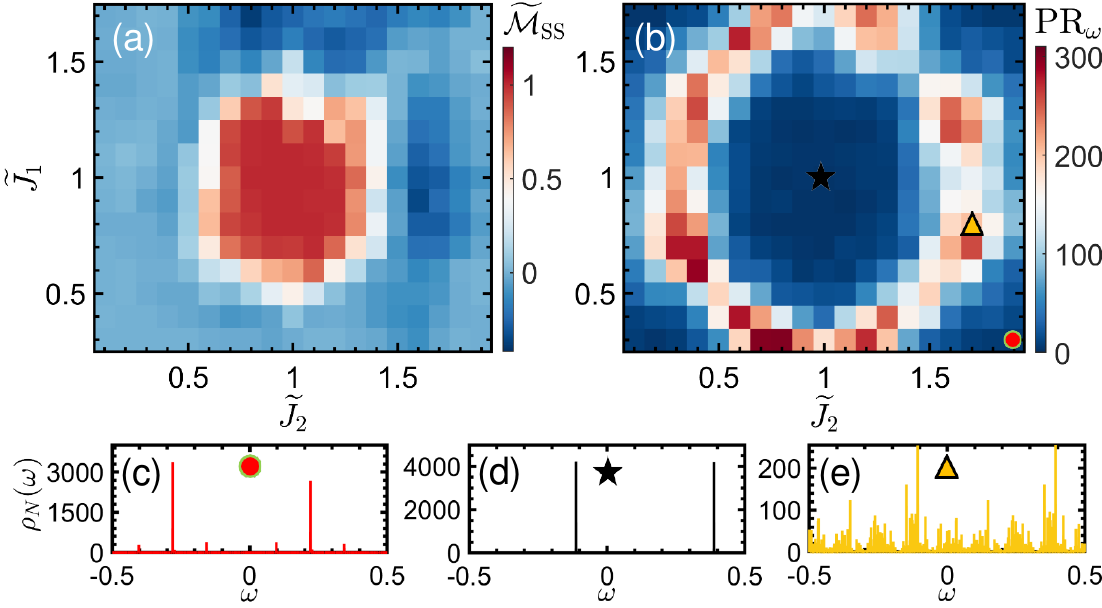}
		\caption{\textbf{Frequency localization in the anomalous Floquet pump model.} (a) Phase diagram for the average $\widetilde{\mathcal M}$ at late time $t/T_3 \in [8\times 10^3, 10^4]$ for the AFPP model with varying coupling constants, where $J_j=\widetilde{J}_j J_j^{(0)}$, where $J_j^{(0)} = J_0 / \sqrt{N_{i,j}}$ for $j\in \{1,2\}$ are the values used previously. (b) Spin-averaged frequency space participation ratio, showing delocalization at the phase boundary. (c-e) Spectral distribution $\rho_N(\omega)$ for 
  (c) initial state $\vert S=0\rangle$, $\widetilde{J}_1=0.3$ and $\widetilde{J}_2=1.9$, (d) initial state $\vert S=0\rangle$, $\widetilde{J}_1=\widetilde{J}_2=1$, and (e) initial state $\vert S=0\rangle$, $\widetilde{J}_1=0.8$ and $\widetilde{J}_2=1.7$. All data use $N=10^4$, and have photon cutoffs $n_{c1,c2}=19$, initial occupation $\mathcal{N}_{i1,i2}=9$, $\omega_2=0.151\sqrt{2}$ and $\omega_1=\omega_2/\sqrt{3}$. 
  } 
		\label{Frequency_localization_AFPP}
\end{figure}

Frequency space delocalization is more easily seen by varying the coupling constants at fixed values of the drive frequencies, which yields a much smoother topological phase diagram by removing issues with frequency commensuration. Therefore, in Fig.~\ref{Frequency_localization_AFPP}, we fix the values of $\omega_{1,2}$ at a point deep in the topological phase and vary the coupling constants $J_{1,2}$. As seen in Fig.~\ref{Frequency_localization_AFPP}(a), the topological phase remains robust within a finite region around the fine-tuned point before transitioning to a topologically trivial phase.

The topological AFPP requires frequency space localization, which can arise from quasi-disorder in frequency space when the frequencies are incommensurate~\cite{Anderson_trnsition_incommensurate, Blekher_Floquet_2level_quasiperiodic, Luck_quasiperiodic_perturbation_2level}. To probe frequency localization, we consider the frequency space participation ratio (PR$_\omega$), which is defined in terms of the spectral distribution $\rho_N (\omega) = \langle \psi_N(\omega) \vert \psi_N (\omega) \rangle$ of the Fourier transform of the full wave function, 
$\vert \psi_N (\omega) \rangle \equiv \frac{1}{\sqrt{\omega_3 N}} \sum_{m=0}^{N-1} e^{-im\omega T_3} \vert \psi (m T_3) \rangle$, 
up to late time $N T_3$. The frequency-space participation ratio, 
$\mathrm{PR}_\omega = \frac{1}{\omega_3} \frac{N}{\int_{-\omega_3/2}^{\omega_3/2} \rho_N(\omega)^2 d\omega}$~\cite{NathanPRLqFTEP}
measures spread over frequency bins. Delocalization results in a more spread-out $\rho_N(\omega)$, whereas localized systems exhibit a finite number of spikes in the $\omega$ spectrum. We choose normalization such that, for large $N \gg 1$, $\mathrm{PR}_\omega$ has finite value when the system is localized and a diverging ($\sim N$) value when it is delocalized. We average $\mathrm{PR}_\omega$ over initial spin states of the qubit. As seen in Figs.~\ref{Frequency_localization_AFPP}(b)-\ref{Frequency_localization_AFPP}(e), the topological phase transition corresponds to a frequency-space delocalization transition~\cite{NathanPRLqFTEP}.


\textbf{\blue{\textit{Experimental perspective.}}}--- While our AFPP model is much simpler than previous topological pumps, it is still vital to ask whether the terms of interest can be realized experimentally on sufficiently fast time scales that dissipation is negligible. While $H_1$ and $H_2$ contain a JC and anti-JC interaction, respectively, we first note that either JC or anti-JC will be sufficient because they can be interchanged by applying a $\pi$-pulse ($X$ gate) before and after the given step. Therefore, our model relies on time-dependent control of either JC or anti-JC interaction, albeit with some tolerance for error due to the topologically protected nature of the phase.

There are two potential avenues we highlight towards realizing the model experimentally, which are illustrated in Fig.~\ref{experimental_AFPP}(a-b). 
The simplest route to realizing a time-dependent JC model is to tune the qubit into resonance with the cavity, requiring strong coupling to avoid dissipation. This could be done by direct tuning of the qubit frequency, e.g., through flux-tuning of the SQUID loop~\cite{Casparis_SQUID_tunable, Luthi_SQUID_tunable} 
or with very strong off-resonant drive by shifting the qubit into resonance using the AC Stark effect. A major concern with this technique is heating and dissipation, which are enhanced by both cavity resonance and fast flux tuning \cite{reed2013thesis}.
\begin{figure}[t]
	\centering
  \includegraphics[width=1\linewidth]{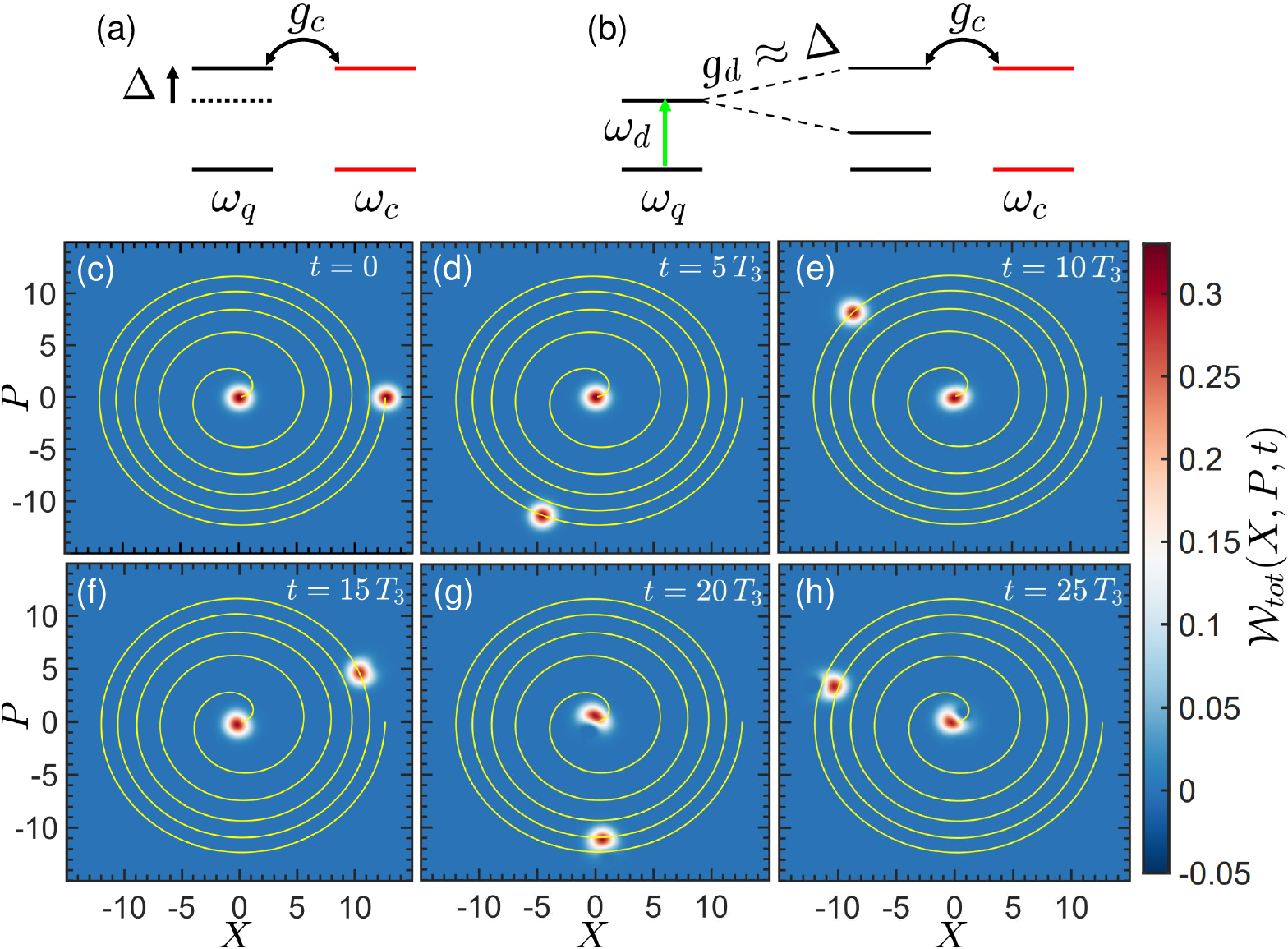}
		\caption{\textbf{Experimental protocols for realizing the AFPP.} (a-b) Two routes to realize JC interactions, as described in the main text. (c-h) Time evolution of the Wigner distribution function $\mathcal{W}_{tot}(X,P,t)$ summed over the $\omega_{1,2}$ modes with the initial state $|0,\alpha=\sqrt{80},0\rangle$ for quadratures $X=(a+a^\dagger)/\sqrt{2}$ and $P=i(a^\dagger - a)/\sqrt{2}$, calculated using QuTiP \cite{JOHANSSONQuTiP1, JOHANSSONQuTiP2}. The peak near the origin represents $\omega_2$ photons. Energy pumping out of the $\omega_1$ mode translates to an in-spiral described by $r^2(t) = r^2(0) - \left(\frac{2}{T_3}\right)t$ and $\theta(t) = - \omega_1 t$ 
  (solid yellow curves), see SM \cite{SM}. Pumping deviates from this topological rate at $t\approx 30T_3$ as the $\omega_1$ wave function approaches the topological phase transition. At the same time, pumping into $\omega_2$ begins, analogous to a chiral edge state turning a corner in the (photon) lattice. All data shown have photon cutoffs $n_{c2}=15$ and $n_{c1}=200$, initial occupation $\mathcal{N}_{i2}=1$ and $\mathcal{N}_{i1}=80$, coupling constants $J_{1}=3/(4\sqrt{\mathcal{N}_{i1}})\approx0.08385$ and $J_{2}=3/(4\sqrt{\mathcal{N}_{i2}})=0.75$, initial spin state $\vert S= 0\rangle$, $\omega_2=0.076\sqrt{2}$ and $\omega_1=\omega_2/\sqrt{3}$. 
        }
		\label{experimental_AFPP}
\end{figure}


A likely better option is to employ a separate resonance by driving the qubit at its bare frequency with an amplitude $g_d \approx \Delta$ that causes one of its Floquet eigenstates to be resonant with the cavity mode. In an appropriate frame of reference, this creates a Jaynes-Cummings interaction directly at first order, meaning that the Jaynes-Cummings interaction rate is set by the same energy $g_c$ as the case of direct qubit-cavity resonance. With realistic parameters, this leads to oscillations on a reasonably fast timescale $\sim 200$ ns (see Supplemental Materials~(SM)~\cite{SM}), which should enable many cycles of topological pumping to occur before qubit or cavity dissipation becomes relevant.


Finally, we ask what observable should be measured to indicate pumping? Bulk ``magnetization'' $\widetilde{\mathcal M}$ is certainly feasible, but as in many related topological phases, the most interesting observable is instead expected to be topological pumping of the edge states. As illustrated in Fig.~\ref{phase_diagram_AFPP}(a), if one of the photons (say $n_2$) is placed near its vacuum state to occupy the edge mode, then the remaining quantized photon gains/loses energy at a rate $W \hbar \omega_1 \omega_3/(2\pi)$
~\cite{Long_Nonadiabatic} where $W$ is the winding number and gain/loss depends on chirality of the bulk phase (which also sets the sign of $W$). A good route to measuring this pumping experimentally will be to prepare the photon $n_1$ in an initial coherent state and measure pumping by its Wigner distribution. The Wigner distribution, a quasi-probability distribution for the two quadratures of the harmonic mode, demonstrates energy pumping through an in-spiral analogous that of a classical harmonic oscillator with constant rate of energy loss. Furthermore, the Wigner distribution is routinely measured in circuit QED and cavity QED experiments \cite{QST_squeeze_microwave_field, Smithey_Wigner_squeezed, Ourjoumtsev_2Photon_Fock}. An illustration of this topological in-spiral is in Fig.~\ref{experimental_AFPP}(c)-(h)
, showing topological pumping that eventually deviates from quantization when the system reaches the topological phase boundary.


\textbf{\blue{\textit{Summary $\&$ outlook.}}}--- We present a detailed protocol designed to achieve non-adiabatic topological photon pumping by leveraging a single qubit coupled to photons in a time-dependent manner, which we call the anomalous Floquet photon pump. This innovative approach can induce a non-trivial topological phase for photons, whether quantized or not, across an extensive range of experimental parameters. The transitions out of this topological phase are primarily driven by delocalization effects in the frequency domain, which significantly alter the system’s behavior. Moreover, we present practical control schemes to realize our protocol using existing experimental setups.

Similar topological pumping features exist in superconducting qubit realizations of the Chern insulator~\cite{Chern_insulator_superconducting_qubit, Floquet_superconduting_qubit}
, but there are key differences that are expected to make the AFPP more robust. First, the Chern pump relies on the preparation of the qubit ground state and adiabatic ramping, whereas the AFPP is insensitive to the initial state and explicitly stabilized by going outside of the adiabatic limit. 
Furthermore, in two spatial dimensions, we found that the anomalous Floquet insulator -- which is in the same class as the AFPP -- is more robust to noise than the Chern insulator~\cite{Timms_QFT_Noise}. 
A key open question for future work will be how the AFPP pumping mechanism is affected by noise, where there is some hope for added stability~\cite{NathanPRLqFTEP}.


\textbf{\blue{\textit{Acknowledgments.}}}---We thank Anushya Chandran, Alicia Kollar, and David Long for valuable discussions. Work was performed with support from the National Science Foundation through award numbers DMR-1945529 and MPS-2228725. Part of this work was performed at the Aspen Center for Physics, which is supported by NSF grant number PHY-1607611. During completion of this manuscript, we became aware of an unrelated work \cite{ChandranEtAlPrepost} in which a similar scheme for dynamical control of Jaynes-Cummings interactions was independently discovered. We used the computational resources of the MRI FASTER by High Performance Research Computing group at Texas A\&M University and the Ganymede and Ganymede2 clusters operated by the University of Texas at Dallas’ Cyberinfrastructure and Research Services Department.

	\bibliography{bib}

\newpage
\begin{widetext}
\section*{Supplementary}
\section{S1. Other models related to Anomalous Floquet qubit pump}\label{s1}
This section derives the Anomalous Floquet Qubit Pump (AFQP) model from a 1D tight-binding model, similar to Ref.~\cite{NathanPRLqFTEP}. First, we consider a chain of fermions with a lattice length of $2L$, incorporating a Floquet drive Hamiltonian with Floquet cycle $T_3=2\pi$ that is equally split into five segments $\tau_1$ to $\tau_5$. Specifically, in the first four segments, we implement the drive Hamiltonian $\hat{H}_{\tau_{1,2,3,4}} =-J \sum_{n=1}^{L}(e^{i\eta_s\lambda}\hat{c}_{2n}^\dag \hat{c}_{2n+d_s}+h.c.)$, where $\eta_{s}=0$ in $\mathcal{\tau}_{1,3}$, $\eta_{s}=-1$ in $\mathcal{\tau}_2$ and $\eta_{s}=1$ in $\mathcal{\tau}_4$. In the last segment $\mathcal{\tau}_5$, we apply an on-site disorder Hamiltonian $\hat{H}_{\tau_{5}}=\sum_{x=1}^{2L}(\frac{\Delta_x}{2})\hat{c}_x^\dag \hat{c}_x$, where $\Delta_x$ refers to the disorder strength. To transform this model into a qubit coupling to two classical-drives model, we group the fermion sites into $L$ units, each comprising sites A and B, and bring the Hamiltonian into the momentum space, where $\vert k, (A, B) \rangle = \frac{1}{\sqrt{L}} \sum_{n=1}^{L} e^{-ink}\vert n, (A, B) \rangle$ with wavenumber $k=\frac{\lambda \pi \rho}{L}$ and $\rho$ being non-negative integers. The resulting momentum representation for the first four segments is $\hat{H}_{\tau_{1,2,3,4}}^{(k)}=-J\sum_{k} (\vert k, B\rangle \langle k, A \vert e^{i\Omega}+h.c.)$, where $\Omega = k$ for $\tau_1$, $\Omega = -\lambda$ for $\tau_2$, $\Omega = 0$ for $\tau_3$, $\Omega = k+\lambda$ for $\tau_4$. For the last segment $\tau_5$, we picked sublattice detuning to be $\Delta_x=2\omega_2$ for $\vert n,A\rangle$ and $\Delta_x=-2\omega_2$ for $\vert n,B\rangle$ so that $\hat{H}_{\tau_5}^{(k)}=\omega_2 \sigma_z$. Given the analogy between the system's states (A and B) in one lattice site unit and the spin states of a fermion, we can view $\vert k, A \rangle$ as spin-up and $\vert k, B \rangle$ as spin-down. The wavenumber $k$ is treated as the phase $ k = \omega_{1}t$ produced by an external driving with frequency $\omega_1$, and the phase $\lambda = \omega_{2} t $ is seen as created by another external drive with frequency $\omega_2$. For one Floquet cycle, the resulting Hamiltonian in each segment, $H_i$, is as follows:
\begin{subequations}
    \begin{align}
    &H_1 = -J(e^{-i\omega_1 t}\sigma^+ +
        e^{i\omega_1 t}\sigma^-) \, ,\\   
    &H_2 = -J(e^{i\omega_2 t}\sigma^+ +
        e^{-i\omega_2 t}\sigma^-) \, ,\\ 
    &H_3 = -J\sigma^x \, ,\\ 
    &H_4 = -J(e^{-i(\omega_1+\omega_2) t}\sigma^+ +
        e^{i(\omega_1+\omega_2) t} \sigma^-)\, , \\ 
    &H_5 = \omega_2\sigma^z \, .
    \end{align}
\end{subequations} 

Although this Hamiltonian has a clear topological phase in the non-adiabatic regime, as $H_4$ requires two drives coupled to the qubit simultaneously, it isn't easy to realize in experiments. To avoid this, we show in the main text that the same topological phase can be obtained by excluding $H_{4,5}$, arriving at the AFQP model, where there are only time segments $\tau_{1,2,3}=T_3/3$ to match with the Hamiltonians $H_{1,2,3}$ accordingly. 

\section{S2. Theoretical differences between Anomalous Floquet qubit pump \& Anomalous Floquet photon pump}\label{s3}
This section derives the differences between the Anomalous Floquet qubit pump (AFQP) and the Anomalous Floquet photon pump (AFPP). Even though AFPP is derived from AFQP, they don't have the same topological properties. Their discrepancies come from the current operators that are defined as $\dot{\Vec{n}}=U^\dag (t,t_0;\Vec{\zeta}_0) (-\nabla_{\zeta}H) U (t,t_0;\Vec{\zeta}_0)$ where $\Vec{\zeta}=\Vec{\omega} t +\Vec{\zeta}_0$ and $\Vec{\omega}=(\omega_1, \omega_2, \omega_3)$. 

In the AFQP model, to facilitate deriving $\dot{\Vec{n}}$ we bring $H_{1,2}$ into the rotating frames $V_{1}=e^{i\omega_1 t/2}\vert 0\rangle \langle 0\vert + h.c.$ and $V_{2}=e^{-i\omega_2 t/2}\vert 0\rangle \langle 0\vert + h.c.$, resulting in the static effective Hamiltonian for $H_j$, denoted $H_{j}^f \equiv V_{j}H_{j}V^\dag +i \dot{V_j} V_j^\dag$. Starting from any time $t=t_0$ in each segment $\tau_i$ with $i\in\{1,2,3\}$, after $\Delta t \leq \tau_i$, we have evolution operator $U_j(t_0+\Delta t)\equiv V^\dag (t_0+\Delta t) e^{-iH_{j}^f \Delta t} V(t_0)$ and the current operator for $\omega_{1,2}$ photon modes given by
\begin{equation}\label{AFQP_n_12}
    \dot{n}_{1,2} = (-\epsilon_{1,2}J)U_0^{\dag}V^{\dag}(t_0)\big(e^{i(H_{1,2}^f)^{\dag}\Delta t} \sigma_y e^{-i(H_{1,2}^f)^{\dag}\Delta t}\big)V(t_0)U_0\, ,
\end{equation}
with $U_0$ as the time evolution operator at the end of the last segment $\tau_{(i-1)}$, $\epsilon_1=-1$ and $\epsilon_2=1$. Notice that the middle term above can be generalized as $e^{i(H_{1,2}^f)^{\dag}\Delta t} \sigma_y e^{-i(H_{1,2}^f)^{\dag}\Delta t} =\big(\sigma_z \sin{\phi_{1,2}} +\sigma_x \cos{\phi_{1,2}} \big) \sin{(2\theta_{1,2} \Delta t)} +\sigma_y \cos{(2\theta_{1,2} \Delta t)}$, 
with
$\sin{\phi_{1,2}} = 2J(4J^2+\omega_{1,2}^2)^{-1/2}, \ 
\cos{\phi_{1,2}} = \epsilon_{1,2}\,\omega_{1,2}(4J^2+\omega_{1,2}^2)^{-1/2}$
and $\theta_{1,2} = \frac{1}{2}(4J^2+\omega_{1,2}^2)^{1/2}$. Assuming the initial state is $\vert S(0)\rangle = \vert 0\rangle$, this allows us to determine $\langle \dot{n}_{2} \rangle$ in segment $\tau_2$ through
\begin{equation}
        \langle \dot{n}_2(T_3/3+\Delta t) \rangle 
        =  {\rm Tr}\Big[-J\vert S(0)\rangle U_0^\dag \Big( \sin{(2\theta_2\Delta t)} \sin{\phi_2} \sigma_z + \big( \big( \cos{\phi_2}\sin{(2\theta_2\Delta t)}-i\cos{(2\theta_2 \Delta t)} \big)\sigma^+ +h.c. \big)\Big)U_0\langle S(0)\vert
         \Big]\,,
\end{equation}
where $U_0 = V_1^\dag (T_3/3) e^{-iH_{1}^f \Delta t} V_1(0)$. Take the beginning of the second time segment $\tau_2$ as an example, when $t=T_3/3$, $\langle \dot{n}_2 (T_3/3)\rangle 
=(-J)\sin{ \frac{ \theta_1 T_3}{3}} \sin{\phi_1}\left(\beta+\beta^*\right)$ with $\beta = e^{i(\omega_2+\omega_1) T_3/3}(i \sin{ \frac{\theta_1 T_1}{3}} \cos{\phi_1}+\cos{ \frac{\theta_1 T_3}{3}} )$.

The discrepancy between AFQP and AFPP in the current operators stems from the differences in time evolution. In what follows, we detail the calculations to address this discrepancy. In the AFPP model, different photon modes are implemented for different segments of the Hamiltonian to couple with the qubit. The photon number operator is denoted as $n_i$ and agrees with the bosonic commutation relation given by $[a , a^\dag]=1$. Utilizing the equation of motion we can calculate the current operators, $\dot{n}_{1,2}^h=i[H^h, n_{1,2}^h]$ and $\Delta n_3^h=U^\dag \Delta n_3^s U$, where the superscript $h$ or $s$ denotes the operators in Heisenberg or Schr\"odinger picture. Accordingly, $n_3$ is defined when the Hamiltonian in the Schr\"odinger picture has sudden changes, i.e., at any start/end points of the time segments, $\Delta n_3^s\equiv\frac{-1}{\omega_3} \Delta H^s$ and $n_{3}^{h} (t=\tau_j) \equiv n_{3}^{h} (t=\tau_j-dt) + \Delta n_3^h$, where $dt$ is a tiny time interval. Writing the time derivative of the average of $n_{1,2}^h$ as $d\langle n_{1,2}^h \rangle/dt$, we have $d\langle n_{1,2}^h \rangle/dt \equiv \langle \psi(t) \vert i[H^s, n_{1,2}^s] \vert \psi(t) \rangle$. Denote $\mathcal{J}_{1,2}= i[H^s, n_{1,2}^s]$, for $0\le t<T_3/3$, we have $\mathcal{J}_{2}= 0$ and $\mathcal{J}_{1} = iJ\bigl( \sigma^+ I_2 a_1^\dag- \sigma^- I_2 a_1 \bigl)$; while for $T_3/3\le t<2T_3/3$, we have $\mathcal{J}_{1}= 0$ and $\mathcal{J}_{2} = iJ \bigl( \sigma^- a_2^\dag I_1 -\sigma^+ a_2 I_1 \bigl)$. 


Since $H_1$ in AFPP is a Jaynes-Cummings (JC) form, and only causes transitions between states $\vert 1, n_1, n_2 \rangle \longleftrightarrow \vert 0,  n_1+1, n_2 \rangle$, it is easy to construct a $2\times2$ block form and derive its eigenvalues and eigenstates. Thus, we arrive at the evolution operator for the first time segment:
\begin{align}
\begin{split}
    e^{\frac{-iH_3\Delta t}{3}} =& \sum_{n_{1}} \bigl(  \vert 0,0,n_2\rangle \langle 0,0,n_2\vert +   e^{-i((n_1+\frac{1}{2})\omega_1 +n_2\omega_2)\Delta t}(\kappa_1 \vert 0,n_1+1, n_2\rangle \langle 0,n_1+1,n_2 \vert\\ &+ \kappa_2 \vert 0,n_1+1, n_2\rangle \langle 1,n_1,n_2 \vert + \kappa_2 \vert 1,n_1,n_2\rangle \langle 0,n_1+1,n_2 \vert + \kappa_1^* \vert 1,n_1,n_2\rangle \langle 1,n_1,n_2 \vert )  \bigl)\,,
\end{split}
\end{align} where $\Omega_1=2J\sqrt{n_1+1}$, $\Delta_1= (\omega_1^2 +\Omega_1^2)^{1/2}$, $\kappa_1=\cos{\frac{\Delta_1 \Delta t}{2}} - i\frac{\omega_1}{\Delta_1} \sin{\frac{\Delta_1 \Delta t}{2}}$ and $\kappa_2= i\frac{\Omega_1}{\Delta_1} \sin{\frac{\Delta_1 \Delta t}{2}}$. Supposedly, if the initial state is $\vert \psi(t=0) \rangle=\vert 0,7,7\rangle$, at time $t=\Delta t$, we obtain $\vert \psi(t) \rangle
= e^{-i((6+\frac{1}{2})\omega_1 +7\omega_2)\Delta t}(\kappa_1 \vert 0,7,7\rangle +\kappa_2\vert 1,6,7 \rangle )$, we have $d\langle n_{2} \rangle/dt =-J\frac{\Omega_1}{\Delta_1} \sin{(\Delta_1 \Delta t)}=0$, which is different from the AFQP model.


\section{S3. Convergence test of Anomalous Floquet photon pump simulation}\label{s2}

In this section, we present the appropriate photon cutoffs for the AFPP simulation. In the main text, we use specific finite cutoffs in the photon spaces, $N_{c1,c2}=19$, and initial photon numbers, $N_{in1,in2}=9$, to sketch out the phase diagrams of AFPP. To verify the convergence of these cutoff values, we compare the variational manifolds at two different cutoffs and calculate the average leakage rate per Floquet cycle at late times. The leakage rate for each Floquet cycle $T_3$ approximately given by 
\begin{equation}
{\rm Leakage\,\, rate\,\, at}\,\, nT_3=\frac{1}{T_3}\big(\vert \langle \psi^{\prime} (nT_3+T_3/3)\vert N_{c1,c2} + 1\rangle \vert ^{2} +\vert \langle \psi^{\prime} (nT_3+2T_3/3)\vert N_{c1,c2} + 1\rangle \vert ^{2} \big)\,,
\end{equation}
where $\vert \psi^{\prime} \rangle $ represents the state resulting from time evolution during the first two time segments $\tau_{1,2}$ of the Floquet cycle starting from a state with cutoff $N_c$ and relaxing the cutoff during time-evolution. Because of the nature of the Hamiltonian, such time evolution only occupies states at $N_c+1$, so this is effectively identical to increasing the cutoff by one.

Since only the topological phases (trivial or non-trivial) remain localized in the bulk, we use a set of parameters designed deep within the topological phase, with $\omega_2=0.076\sqrt{2}$ and $\omega_1=\omega_2/\sqrt{3}$, to perform the convergence test. Figure~\ref{sf1} shows the leakage rates at late times, $t\in [8\times 10^3,10^4]\,T_3$, for both spin-up (blue marker) and spin-down (red marker) initial states. As shown in Fig.~\ref{sf1} with cutoffs $N_{c1,c2}=19$, both cases exhibit consistently low leakage rates and demonstrate good convergence. The average late-time leakage for the spin-up initial state is approximately $3.595225240561944\times 10^{-13}$ and for the spin-down initial state, it is around $6.190703299893656\times 10^{-11}$.
\begin{figure*}[t]
		\centering
  \includegraphics[width=0.8\linewidth]{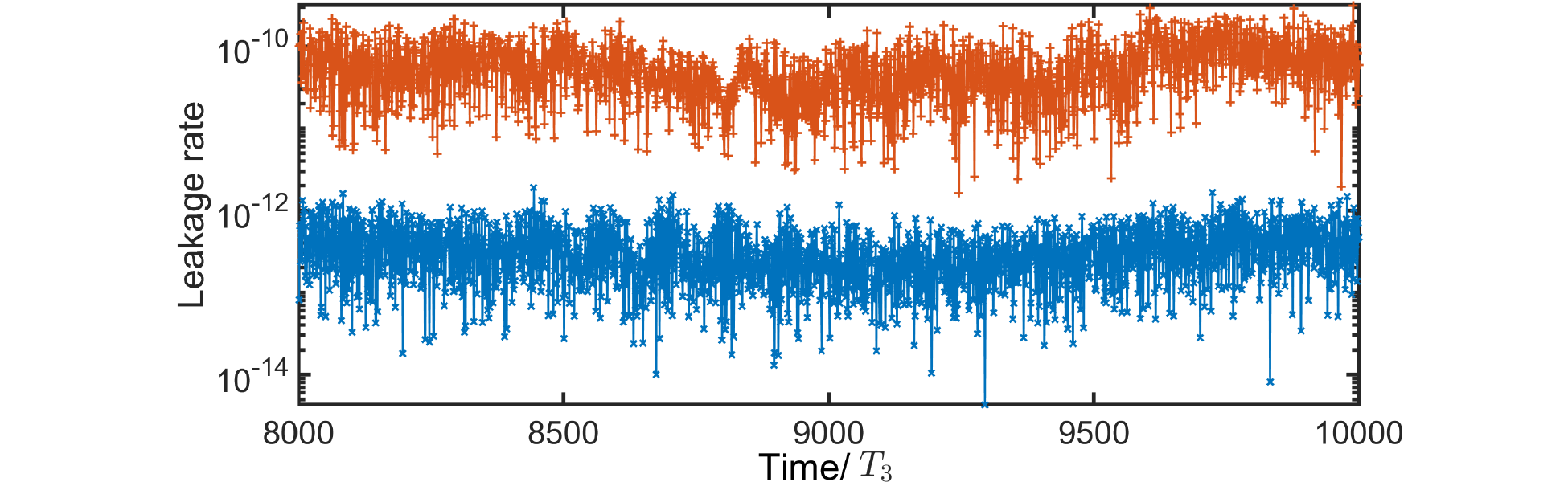}
    \caption{\textbf{Convergence test of Anomalous Floquet photon pump simulation}.  Leakage rate for parameters $\omega_2=0.076\sqrt{2}$, $\omega_1=\omega_2/\sqrt{3}$ with photon cutoffs $N_{c1,c2}=19$. The red line corresponds to the spin-down initial state, and the blue line represents the spin-up initial state. }
		\label{sf1}
\end{figure*}

\section{S4. Time-dependent Jaynes-Cummings via Rabi resonance}
This section analyzes how to get strong, tunable JC coupling using resonance between dressed-driven qubit states and a cavity mode. Let the drive to be resonant with the qubit, the Hamiltonian is similar to Eq.\eqref{eq_2photon_drive}:
\begin{equation}\label{eq_resonate}
    H_{exp_1}(t)=-\frac{\omega_{q}}{2}Z+\omega_{c}a^{\dagger}a-g_{d}\cos\left(\omega_{q}t\right)X-g_{c}\left(a+a^{\dagger}\right)X\,.
\end{equation}

Next, we enforce another resonance condition, namely $g_{d}=\omega_{c}-\omega_{q}$, which comes from the fact that the resonantly driven qubit has energy levels $\omega_{q}\pm g_{d}$ and one of these levels is resonant with the cavity under this assumption. To see how this works, start by going to the general rotating frame for resonant Rabi drive $V_{exp_1}^{(1)}=\exp\left[i\omega_{q}\left(-\frac{1}{2}Z+a^{\dagger}a\right)t\right]$, the effective Hamiltonian is 
\begin{equation}
    H_{exp_1}^{\prime}=\left(\omega_{c}-\omega_{q}\right)a^{\dagger}a-g_{d}\cos\left(\omega_{q}t\right)\left[X\cos\left(\omega_{q}t\right)+Y\sin\left(\omega_{q}t\right)\right]-g_{c}\left(ae^{-i\omega_{q}t}+a^{\dagger}e^{i\omega_{q}t}\right)\left[X\cos\left(\omega_{q}t\right)+Y\sin\left(\omega_{q}t\right)\right]\,.
\end{equation}

Under the usual assumption that $\omega_{q}\gg\left|\omega_{c}-\omega_{q}\right|=g_{d}\gg g_{c}$, we have terms oscillating at $2\omega_{q}$ and others that are static. Let's average out the oscillating ones to get a time-averaged Hamiltonian:
\begin{equation}
    \bar{H}_{exp_1}^{\prime}=\left(\omega_{c}-\omega_{q}\right)a^{\dagger}a-\frac{g_{d}}{2}X-\frac{g_{c}}{2}\left(a\left(X-iY\right)+a^{\dagger}\left(X+iY\right)\right)\,.
\end{equation}

Now we make two further assumptions – first that the drive strength is resonant ($g_{d}=\omega_{c}-\omega_{q}$), and second that the cavity coupling is weak compared to the drive ($g_{d}\gg g_{c}$). In that case, we have 
\begin{equation}
\bar{H}_{exp_1}^{\prime}=g_{d}\left(a^{\dagger}a-\frac{X}{2}\right)-\frac{g_{c}}{2}\left(a+a^{\dagger}\right)X-\frac{g_{c}}{2}i\left(a^{\dagger}-a\right)Y\,,
\end{equation}
where the last term is weak. Therefore, we can do a second rotating frame transformation $V_{exp_1}^{(2)}=\exp\left[ig_{d}\left(-\frac{1}{2}X+a^{\dagger}a\right)t\right]$, the effective Hamiltonian is
\begin{equation}
    H_{exp_1}^{\prime \prime}=-\frac{g_{c}}{2}\left(ae^{-ig_{d}t}\left(X-i\left(Y\cos\left(g_{d}t\right)+Z\sin\left(g_{d}t\right)\right)\right)+h.c.\right)\,.
\end{equation}

Again, dropping fast-oscillating terms at frequency $g_{d}$ and $2g_{d}$, we obtain $\bar{H}_{exp_1}^{\prime\prime}=\frac{g_{c}}{2}\left(a\sigma_{x}^{-}+a^{\dagger}\sigma_{x}^{+}\right)$, where $\sigma_{x}^{\pm}=\frac{1}{2}\left(Z\mp iY\right)$ are the x-basis raising/lowering operators functioning like $\sigma^{+}=\left(\begin{array}{cc}
0 & 1\\
0 & 0
\end{array}\right)$ and $\sigma^{-}=\left(\begin{array}{cc}
0 & 0\\
1 & 0
\end{array}\right)$ in the z-basis. In Figs.~\ref{sf2}(b) and~\ref{sf2}(c), we show two simulation examples that realize this model. 
\begin{figure*}[t]
		\centering
  \includegraphics[width=1\linewidth]
  {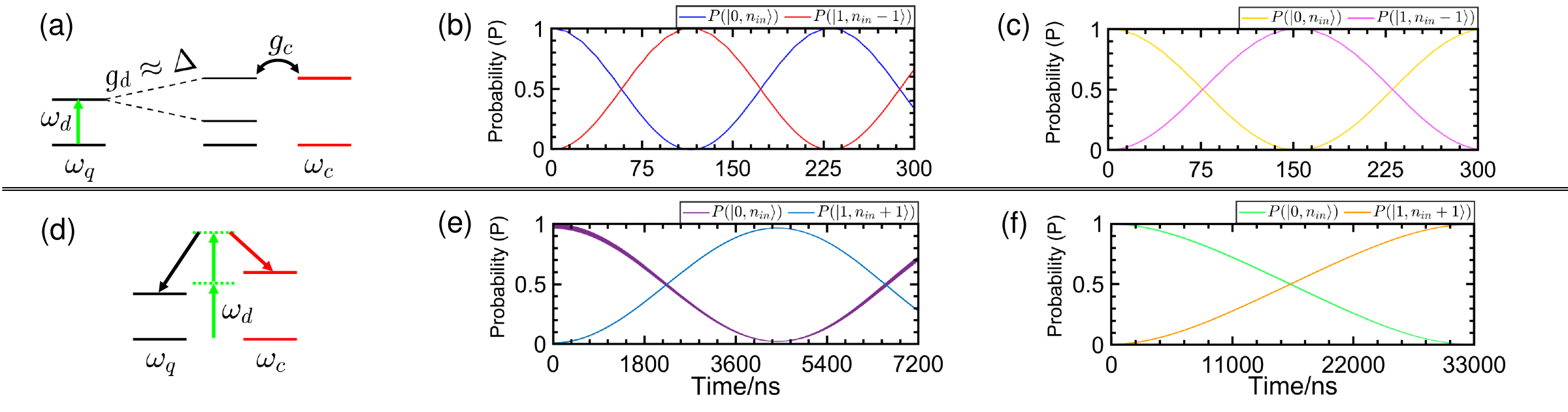}
    \caption{\textbf{Tunable Jaynes-Cummings and anti-Jaynes-Cummings simulations}. (a) Route to realize JC interaction via Rabi resonance, as described in the main text. (b-c) The Jaynes-Cummings~(JC) experimental simulation for initial photon number $ n_{in}=2$, cutoff $n_c=5$, cavity frequency $\omega_{c}=4.3$ GHz, with (b) qubit frequency $\omega_{q}=3.8$ GHz, coupling constant $g_{c}=0.02$; with (c) qubit frequency $\omega_{q}=3.75$ GHz, coupling constant $g_{c}=0.015$. (d) Route to realize anti-JC interaction via 2-photon drive. (e-f) The anti-Jaynes-Cummings~(JC) experimental simulation for initial photon number $ n_{in}=2$, cutoff $n_c=5$, cavity frequency $\omega_{c}=10$ GHz, qubit frequency $\omega_{q}=11$ GHz, coupling constant $g_{c}=0.02$, with (e) classical drive strength $g_{d}=0.05$ and frequency $\omega_{d}=10.5063103368895$ GHz; with (f) classical drive strength $g_{d}=0.02$ and frequency $\omega_{d}=10.5020813368895$ GHz. 
    }
		\label{sf2}
\end{figure*}
\section{S5. Time-dependent anti-Jaynes-Cummings via 2-photon drive}\label{S5}
This section examines a simplified Hamiltonian confined to the qubit manifold with a conventional drive, where tunable anti-JC coupling arises as a third-order process. Starting from the general form of circuit QED Hamiltonian 
\begin{equation}\label{eq_2photon_drive}
    H_{exp_2}(t)=-\frac{\omega_{q}}{2}Z+\omega_{c}a^{\dagger}a+2g_{d}\cos\left(\omega_{d}t\right)X+g_{c}\left(a+a^{\dagger}\right)X\,,
\end{equation}
the key idea is to set $\omega_{d}\approx(\omega_{q}+\omega_{c})/2$, allowing two photons from the classical drive to be converted into two excitations in the qubit and cavity modes. 
To mathematically prove this, we will derive the effective Hamiltonian using operator perturbation theory. First, we transform the Hamiltonian into a rotating frame with respect to the qubit and cavity. Specifically, the rotating frame operator is given by $V(t)=e^{-i\left(\omega_{d}+\Delta^{\prime}\right)Zt/2}e^{i\left(\omega_{d}-\Delta^{\prime}\right)a^{\dagger}at}$, where $Z$ is the Pauli-$Z$ operator for the qubit, and $a^\dagger a$ is the photon number operator for the cavity mode. The rotating-frame effective Hamiltonian is the given by transforming the original Hamiltonian $H$ into this rotating frame:
\begin{align}
    \begin{split}\label{H_exp_prime}
    H_{exp_2}^{\prime} =&-\frac{\omega_{q}-\left(\omega_{d}+\Delta^{\prime}\right)}{2}Z+\left[\omega_{c}-\left(\omega_{d}-\Delta^{\prime}\right)\right]a^{\dagger}a+2g_{d}\cos\left(\omega_{d}t\right)\left[\sigma^{+}e^{-i\left(\omega_{d}+\Delta^{\prime}\right)t}+h.c.\right] \\ &+g_{c}\left(ae^{-i\left(\omega_{d}-\Delta\right)t}+h.c.\right)\left[\sigma^{+}e^{-i\left(\omega_{d}+\Delta^{\prime}\right)t}+h.c.\right].
\end{split}
\end{align}
This rotating-frame Hamiltonian will allow us to isolate the interactions relevant for deriving the effective anti-JC coupling.

In the previous discussion, we set $\omega_{d}\approx\left(\omega_{q}+\omega_{c}\right)/2$, so a natural choice for $\Delta^{\prime}$ would be the one that causes the first two terms in Eq.~\eqref{H_exp_prime} to vanish. However, let us assume some additional detuning, implying that the two-photon resonance for $\omega_{d}$ is close but not perfectly exact. Specifically, we define the detunings as:\begin{equation}
    \delta_{q} \equiv\omega_{q}-\left(\omega_{d}+\Delta^{\prime}\right)\ \text{and}\ 
    \delta_{c} \equiv\omega_{c}-\left(\omega_{d}-\Delta^{\prime}\right)=0\,,
\end{equation}
which give $\omega_{d}=\omega_{c}+\Delta^{\prime}$ and $\delta_{q}\equiv \omega_{q}-\omega_{c}-2\Delta^{\prime}\approx 0$.
If we drive the system exactly at resonance, both detunings would need to vanish, resulting in the condition $2\Delta^{\prime}=\omega_{q}-\omega_{c}$.

Next, we assume the following hierarchy of energy/frequency scales: the bare photon and qubit frequencies are very large. The splitting between them, $\sim\left|\Delta^{\prime}\right|$, is somewhat large but still much smaller than these frequencies. The drive strengths ($g_{c,d}$) and detunings ($\delta_{c,q}$) are small. Therefore, these scales satisfy $\omega_{q,c,d}\gg\left|\Delta^{\prime}\right|\gg g_{c,d},\delta_{c,q}$, where $\omega\gg\left|\Delta^{\prime}\right|$ can be relaxed and gives rise to some other additional terms. Under these assumptions, we can neglect the fast rotating terms $\sim e^{\pm2i\omega_{d}t}$ in Eq.~\eqref{H_exp_prime} and simplify it to: $H_{exp_2}^{\prime\prime}\approx-\frac{\delta_{q}}{2}Z+\delta_{c}a^{\dagger}a+g_{d}(\sigma^{+}e^{-i\Delta^{\prime} t}+h.c.)+g_{c}(a\sigma^{-}e^{2i\Delta^{\prime} t}+h.c.)$.

With these manipulations, we have transformed the original problem into a Floquet problem with a fast oscillating frequency $\Delta^{\prime}$. Since all coefficients in the Hamiltonian are small, we can apply the van Vleck high-frequency expansion~\cite{PhysRevX.4.031027} to derive an effective Hamiltonian: 
\begin{equation}
    H_{eff}	=  H^{(0)}+\frac{1}{2}\sum_{n\neq0}\frac{\left[H^{(n)},H^{(-n)}\right]}{n\Delta^{\prime}}+\frac{1}{2}\sum_{n\neq0}\frac{\left[\left[H^{(n)},H^{(0)}\right],H^{(-n)}\right]}{n^{2}{\Delta^{\prime}}^{2}} +\frac{1}{3}\sum_{m\neq0}\sum_{n\neq0,m}\frac{\left[\left[H^{(m)},H^{(n-m)}\right],H^{(-n)}\right]}{mn{\Delta^{\prime}}^{2}}+O({\Delta^{\prime}}^{-3})\,.
\end{equation}

From the simplified rotating-frame effective Hamiltonian $H_{exp_1}^{\prime \prime}$, we obtain five non-vanishing Fourier harmonics $H^{(0)} =-\frac{\delta_{q}}{2}Z+\delta_{c}a^{\dagger}a$, $H^{(1)}=\left(H^{(-1)}\right)^{\dagger}=g_{d}\sigma^{-}$, and $H^{(2)}=\left(H^{(-2)}\right)^{\dagger}=g_{c}\sigma^{-}a$. Here, we can derive the dispersive/Stark (first order) correction:
\begin{equation}
    H_{eff}^{[1]}	=\frac{1}{2}\sum_{n\neq0}\frac{\left[H^{(n)},H^{(-n)}\right]}{n\Delta^{\prime}}=\left(-\frac{g_{d}^{2}}{\Delta^{\prime}}-\frac{g_{c}^{2}}{4\Delta^{\prime}}-\frac{g_{c}^{2}}{2\Delta^{\prime}}a^{\dagger}a\right)Z+\frac{g_{c}^{2}}{4\Delta^{\prime}}\,.
\end{equation}
This represents the AC Stark shift from $g_{d}^{2}$ and the dispersive shift of qubit and cavity frequencies from $g_{c}^{2}$. The second order correction includes the anti-JC term that we're after:
\begin{equation}
    H_{eff}^{[2]} =\frac{1}{2}\sum_{n\neq0}\frac{\left[\left[H^{(n)},H^{(0)}\right],H^{(-n)}\right]}{n^{2}{\Delta^{\prime}}^{2}}+\frac{1}{3}\sum_{m\neq0}\sum_{n\neq0,m}\frac{\left[\left[H^{(m)},H^{(n-m)}\right],H^{(-n)}\right]}{mn{\Delta^{\prime}}^{2}}\,\\
    =-\frac{g_{c}g_{d}^{2}}{{\Delta^{\prime}}^{2}}\left(\sigma^{+}a+\sigma^{-}a^{\dag}\right)\,.
\end{equation}
Combining the derivations for both correction terms, the resulting effective Hamiltonian is
\begin{equation}
H_{eff}\approx\left(-\frac{\delta_{q}}{2}-\frac{g_{d}^{2}}{\Delta^{\prime}}-\frac{g_{c}^{2}}{4\Delta^{\prime}}\right)Z-\frac{g_{c}^{2}}{2\Delta^{\prime}}a^{\dagger}aZ+\delta_{c}a^{\dagger}a+\frac{g_{c}^{2}}{4\Delta^{\prime}}-\frac{g_{c}g_{d}^{2}}{{\Delta^{\prime}}^{2}}\left(\sigma^{+}a+\sigma^{-}a^{\dag}\right)\,.
\end{equation}
In order to “activate” the aJC term, we need the states $|0,n\rangle$  and $|1,n+1\rangle$ to be resonant, i.e.,
\begin{equation}
     \left(-\frac{\delta_{q}}{2}-\frac{g_{d}^{2}}{\Delta^{\prime}}-\frac{g_{c}^{2}}{4\Delta^{\prime}}\right)+\left(\delta_{c}-\frac{g_{c}^{2}}{2\Delta^{\prime}}\right)n	=-\left(-\frac{\delta_{q}}{2}-\frac{g_{d}^{2}}{\Delta^{\prime}}-\frac{g_{c}^{2}}{4\Delta^{\prime}}\right)+\left(\delta_{c}+\frac{g_{c}^{2}}{2\Delta^{\prime}}\right)\left(n+1\right)\,.
\end{equation}

Achieving this resonance will indeed require a delicate balance due to the $n$-dependent shift. However, perfect precision might not be critical because response functions are topologically protected. To minimize the impact of the $n$-dependence relative to JC coupling, it will be useful to have $g_{d}$ significantly stronger than $g_{c}$. Solving this equation, we present 
simulations corresponding to two different sets of parameters in Figs.~\ref{sf2}(e) and~\ref{sf2}(f). 
\section{S6. Energy pump rate}
In this section we calculate the energy pump rate and consequently, the rate of radial distance change in the Wigner distribution function (WDF) calculations. 
In our numerical calculations of WDF to obtain the energy loss rates out of mode $\omega_3$, we used a coherent state of the form $\ket{\alpha} = e^{-|\alpha|^2/2}\sum_{n=0}^\infty \frac{\alpha^n}{\sqrt{n !}}\ket{n}$ for $\alpha$ being a complex number and the eigenvalue of annihilation operator and $\ket{n}$ is the Fock state of photons. For such a state, the average energy of the photonic mode is given by (excluding the zero point energy)
\begin{equation}
    E_i = \omega\bra{\alpha} a^\dagger a \ket{\alpha} = \omega|\alpha|^2\, .
\end{equation}
A standard picture of a single-mode electromagnetic field is to represent it in terms of the harmonic oscillator of unit mass given by $H = \frac{1}{2}\left( \hat p^2 + \omega^2 \hat x^2 \right)$, where $\omega$ is the frequency of the mode and $x$ and $p$ are canonical position and momentum related to the electric and magnetic field. Upon quantization of the field in terms of the raising and lowering operators, we can write $\hat x = \frac{1}{\sqrt{2\omega}}\left( \hat a^\dagger + \hat a \right)$ and $\hat p = i\sqrt{\frac{\omega}{2}}\left( \hat a^\dagger - \hat a \right)$, given the choice of the coherent state for the initial state, we can find the expectation values of $\hat x$ and $\hat p$ using
\begin{equation}
\braket{\hat x} = \frac{1}{\sqrt{2\omega}}\left( \alpha^* + \alpha \right), \quad \braket{\hat p} = i\sqrt{\frac{\omega}{2}}\left( \alpha^* - \alpha \right)\,.
\end{equation}
By defining $X =  x \sqrt{\omega}$ and $P =  p /\sqrt{\omega}$ we can make use of them as two classical variables independent of $\omega$, and define Wigner space axis by mapping $\hat x \rightarrow X=\hat x\sqrt{\omega}$ and $\hat p \rightarrow P=\hat p/\sqrt{\omega}$.
In this Wigner space, the peak of WDF along $X$ and $P$ axis for a coherent is given by $\left( \alpha^* + \alpha \right)/\sqrt{2}$ and $i\left( \alpha^* - \alpha \right)/\sqrt{2}$, respectively. 
Therefore the relation between energy and radial distance of the Wigner function distribution peak from the origin, $r$, is given by
\begin{equation}
     \frac{r^2}{2}= \frac{1}{2} \left( X^2 + P^2 \right) =\frac{1}{2} \left( \left( \frac{\left( \alpha^* + \alpha \right)}{\sqrt{2}} 
 \right)^2 + \left(i \frac{\left( \alpha^* - \alpha \right)}{\sqrt{2}} 
 \right)^2 \right) = |\alpha|^2 \quad \rightarrow \frac{E_i}{\omega} = \frac{r^2}{2}
\end{equation}
On the other hand, the energy current out of one mode in our model is $d{E}_i/dt=-\omega_i/T_3 $, therefore, we can write 
\begin{equation}
    \frac{d{E}_i}{dt}= \omega_ir\frac{dr}{dt}=-\frac{\omega_i}{T_3} \longrightarrow \frac{dr^2}{dt}=-\frac{2}{T_3 } \longrightarrow r^2(t) = r^2(0) -\frac{2}{T_3}t\, ,
\end{equation}
where $r(0)$ and $r(t)$ are the radial distance at initial time $t=0$ and later time $t$, respectively. This equation shows that the path the Wigner distribution follows while pumping energy is the same irrespective of the cavity mode frequency ($\omega_i$). Only how fast it moves on that path is defined by $\theta(t) = -\omega_i t$.

\end{widetext}
\end{document}